\documentclass[doublecol]{epl2} 
\usepackage{amsmath}
\def\be{\begin{equation}}
\def\ee{\end{equation}}

\def\bc{\begin{center}} 
\def\ec{\end{center}}
\def\bea{\begin{eqnarray}}
\def\eea{\end{eqnarray}}

\title{Topology-Induced Inverse Phase Transitions}
\shorttitle{Topology-Induced Inverse Phase Transitions} 

\author{D.De Martino\inst{1}, S.Bradde\inst{2}, L.Dall'Asta\inst{3} \and M.Marsili\inst{4}}

\institute{                    
\inst{1} Dipartimento di Fisica, Sapienza Universit\'{a} di Roma, P.le A.Moro 2, 00815, Rome, Italy\\
\inst{2} MSKCC-Computational Biology Center, 1275 York Avenue, New York, USA \\
\inst{3} Politecnico di Torino, Corso Duca degli Abruzzi 24, 10129 Torino, Italy\\
\inst{4} The Abdus Salam International Center for Theoretical Physics, Strada Costiera 11, 34014, Trieste, Italy

}

\abstract{ Inverse phase transitions are striking phenomena in which an apparently more ordered state disorders under cooling. This behavior can naturally emerge in tricritical systems on
heterogeneous networks and it is strongly enhanced by the presence of disassortative degree correlations. We show it both analytically and numerically, providing also a microscopic interpretation
of inverse transitions in terms of freezing of sparse subgraphs and coupling renormalization.}

\begin{document}

\maketitle

\section{Introduction}
Network-based representations are appropriate tools to
describe many real-world systems and to classify their
structural and functional properties using few 
topological measures, such as for instance the degree distribution,
degree correlations and clustering\cite{1}.
Networks characterized by degree distributions with power-law tails
have attracted a lot of attention, mostly because of their unusual
structural and dynamical properties  \cite{2,3,3b}.
However these studies are mostly limited to uncorrelated 
random graphs, i.e. graphs characterized solely by their degree distribution.
Many real-world networks show instead
degree correlations. For instance, social networks are usually characterized by positive degree correlations (assortative mixing),
because high-degree nodes tend to be directly connected and to
form cliques \cite{4}. In technological and biological networks,
instead, the hubs are preferentially connected to low 
degree nodes (disassortative mixing)\cite{5}. 
Degree correlations can strongly affect the overall behavior of dynamical processes, 
as it was recently shown for percolation\cite{6} and diffusion processes\cite{7}. 

In this Letter, we show that some amount of degree heterogeneity and
negative degree correlations can be responsible for the 
occurrence of inverse phase transitions. The latter are counterintuitive 
phenomena in which a system, starting from a high 
temperature disordered phase first undergoes a phase transition
to a more ordered phase than comes back to a disordered
one as a result of monotonic decrease of temperature.
Examples of this curious behavior are observed in a variety of systems, such as liquid
binary mixtures, $^3$He-$^4$He isotopes, ultra-thin films, 
liquid crystals, disordered high-T superconductors and 
polymeric solutions \cite{8}. An inverse transition implies the inversion of the standard ratio between the entropic 
content of the two phases\cite{9}. This behavior has a simple
microscopic explanation for some systems, such as water
solutions of methyl-cellulose polymers, in which the more
interacting unfolded state is entropically favored because
it admits many more microscopic configurations than the
non-interacting folded one. As suggested by the Flory-
Huggins theory of polymer melts\cite{10}, a general way of
triggering an inverse phase transition is to introduce a
temperature-dependent interaction. This latter naturally
emerges in spin systems with higher degeneracy of interacting 
states\cite{11}. 
Moreover, an inverse freezing transition between glassy and paramagnetic phases 
exists in tricritical spin glasses\cite{12},\cite{13},
\cite{13b} and in spin-glasses on small world graphs\cite{13c}. 

Here we study the tricritical Ising model defined on random graphs,  
providing evidence of a novel topological mechanism 
that is responsible of an inverse melting transition. 
Using the cavity method,  we show that degree fluctuations can 
generate a reentrance in the low temperature region of
the phase diagram. The physical reason at the origin of this behavior 
is identified in the partial freezing of the microscopic degrees of freedom that generates a temperature-dependent renormalization of 
the effective interaction couplings. The mechanism, that requires some amount of degree heterogeneity to occur,  is amplified  by disassortative mixing. 
This is of great relevance because many real-world technological and biological 
networks are disassortative and degree anti-correlations have been recently shown
to be a natural feature of heterogeneous networks\cite{14}.

\section{The Model}
The tricritical Ising model, known as
Blume-Capel (BC) model\cite{15}, describes spin-1 variables
$s_i= \{0, \pm1\}$ defined on the sites of a lattice or a graph,
and interacting according to the Hamiltonian
\begin{equation}\label{bc}
H_{BC} = - J \sum_{\langle i,j\rangle} s_i s_j + \Delta \sum_{i} s_i^2
\end{equation}
where the first sum runs over nearest neighbor pairs $\langle i,j \rangle$
and the chemical potential $\Delta$ controls the density.
It is well known that the BC model exhibits a rich $T-\Delta$ phase diagram
with lines of first-order and second-order phase transitions
dividing the ferromagnetic phase at low temperature $T$ and chemical potential $\Delta$ from the paramagnetic phase at large $T$ and/or $\Delta$ 
\cite{15}. The composition of the paramagnetic phase changes gradually
and continuously from a purely disordered one (spins $\pm1$) at large $T$ and low $\Delta$, to a vacancy-dominated state for large $\Delta$ and low $T$.
In the following, we consider spins defined on the nodes of an uncorrelated random graph with degree sequence ${k_1,k_2, \dots, k_N}$ randomly drawn
from a given degree probability distribution $P(k)$.


\begin{figure}
\onefigure[width=0.45\textwidth]{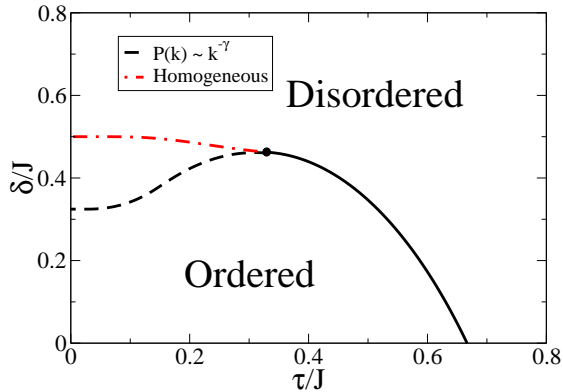}
\caption{Phase diagram $(\tau,\delta)$ of the Blume-Capel model in the Curie-Weiss approximation for random graphs. 
The second order continuous branch has the universal scaling form of the Eq.\ref{rescaling}. The first order discontinuous branches 
are different for homogeneous (random regular graph of degree $K=5$) and heterogeneous graphs (random graph with power-law degree distribution $P(k) \propto k^{-\gamma}$, with $\gamma = 3.5$) respectively.}
\label{fig.1}
\end{figure}

\section{Role of Degree Heterogeneity}
A qualitative insight on the role played by degree heterogeneity in determining the critical behavior of the BC model can be obtained within a Curie-Weiss (CW) approximation.
This is comparable to an ``annealed" or a ``random neighbor'' approximation where the adjacency
matrix $a_{i,j}$\footnote{$a_{i,j}$ is $1$ if $i$ and $j$ are neighbors,
$0$ otherwise.} is replaced in the Hamiltonian by the probability $k_i k_j /\langle k \rangle N$ that the link between nodes $i$ and $j$ is present \cite{2}. 
This approximation allows the factorization of the free energy and the analytic computation of the partition function by Gaussian integration. 
It follows that the magnetization  $\mu$ of a neighbor of a randomly chosen site satisfies
the mean-field self-consistent equation $\mu = F(\mu)$ where
\begin{equation}\label{fmu}
F(\mu) = \sum_{k=k_{min}}^{k_{max}} \frac{k P(k)}{\langle k \rangle} \frac{\sinh(\beta k \mu)}{e^{\beta \Delta}/2 + \cosh(\beta k \mu)}.
\end{equation}

This equation has a paramagnetic solution $\mu=0$ whose stability is determined imposing $F'(0)\le 1$. 
Following Landau theory, the equation $F'(0)=1$ defines a curve of second-order critical points 
between the paramagnetic and the ferromagnetic phases (the $\lambda$-line) 
till, upon decreasing the temperature, 
the third-order term in the free energy\footnote{This can be obtained by integrating eq.\ref{fmu}.} expansion vanishes $F'''(0) = 0$ 
(the tricritical point), then the transition becomes first order and the aforementioned condition
gives us only the location of the spinodal point beyond which the paramagnetic solution is not stable.  
In terms of the rescaled variables 
$\delta = \Delta \langle k \rangle /\langle k^2 \rangle$ and  $\tau = T \langle k \rangle /\langle k^2 \rangle$ 
the $\lambda$-line has the form
\begin{equation}\label{rescaling}
\delta = \tau \log [2(1/\tau-1)],
\end{equation}
whereas the tricritical point is $\tau_c = \frac{1}{3}$, $\delta_c =\frac{\log4}{3}$, 
that coincides with the maximum of the curve Eq.\ref{rescaling}.
Interestingly, the obtained mean-field $\lambda$-line depends on the graph properties only
through the ratio $\langle k^2 \rangle /\langle k \rangle$, which determines the scaling factor.

Lowering the temperature beyond the tricritical point $(\tau_c, \delta_c)$, we find a different dependence of the (first-order) transition line on the topological properties. 
It is useful to look directly to the $T=0$ limit, where Eq. \ref{fmu}
becomes 
\begin{equation}
\mu = \langle k \Theta(k \mu -\Delta)\rangle /\langle k \rangle
\end{equation}
and $\Theta(x)$ is the Heaviside step function. A direct calculation, done approximating sums on $k$ with integrals, reveals the following behavior.
For graphs with a degree distribution which falls off faster than $k^{-3}$
for large $k$, we find that the ferromagnetic state with $\mu=1$ is possible at $T=0$ when $\Delta<k_{min}$. In order to access the stability property of this solution
one needs to compare its energy $E[\mu] = -\langle k\rangle \mu^2/2+\Delta \mu$ with the paramagnetic energy ($E=0$) of the state with $s_i=0$ $\forall i$. 
This shows that the $\mu=1$ state is stable as long as $\Delta < \langle k \rangle /2$. So for graphs with $k_{min}>\langle k \rangle/2$, 
as for example graphs with $P(k) \simeq k^{-\gamma}$ and $\gamma>3$, the coexistence region at $T=0$ extends from $0 \leq \Delta \leq k_{min}$ with a first order phase transition at $\langle k \rangle/2$. 
For scale free graphs with $2 \leq \gamma \leq 3$ the same arguments show that, at  $T=0$, a ferromagnetic state with $\mu\sim \Delta^{-\frac{3-\gamma}{\gamma-2}}$ exists for $\Delta <k_{max}$
and it is thermodynamically stable for $\Delta<(\langle k \rangle/2)^{\gamma-2}$. The same holds for $\gamma<2$ with the important difference that $\langle k \rangle$ now diverges with $k_{max}$ and that the magnetization
$\mu$ is only weakly dependent on $\Delta$. 

Some of the results are summarized in Fig.\ref{fig.1} which displays the phase diagram obtained in the CW approximation 
for a homogeneous graph (regular random graph with degree $K=5$) and for a heterogeneous graph with $P(k) \propto k^{-3.5}$. 
The first-order branch was determined by equating 
the free energy of the paramagnetic and the ferromagnetic solutions.
While the lines of continuous phase transitions collapse using the rescaled variables $(\tau,\delta)$, 
the behaviors of the discontinuous transitions remain different as they depends on the average degree $\langle k \rangle$, and not on the ratio $\langle k^2\rangle/\langle k\rangle$. The reentrant behavior of the first-order transition line is  evident  in the case of heterogeneous graphs.   
A similar result holds for queuing models on random graphs, 
where the transition point that divides congested and free phases 
scales differently with the moments of the degree distribution 
for continuous and discontinuous transitions, 
respectively\cite{congestion}. 
In summary, the CW approximation would suggest that in the BC model an inverse phase transition can be triggered
by a different scaling of first and second order critical lines with the moments of the degree distribution. We see in the following that, although the role of degree heterogeneity is
crucial to have a reentrance, this is not strictly associated to the different scaling of the first-order and second-order phase transitions.


\begin{figure}
\onefigure[width=0.49\textwidth]{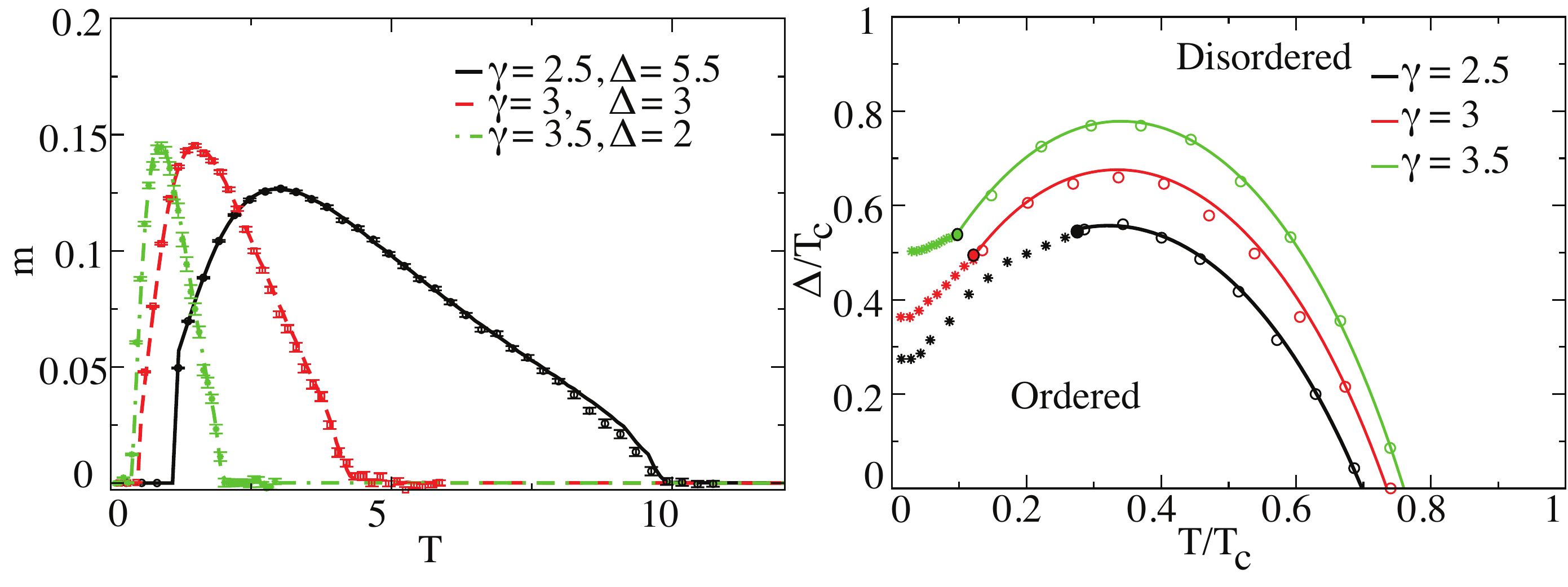}
\caption{Reentrant behavior in a random graph of size $N = 10^4$ and
power-law degree distribution $P(k) \propto k^{-\gamma}$.
Left) Curves $m(T)$ for different values of 
 $\gamma$ and $\Delta$. Right) Phase diagram for 
 $\gamma= 2.5,3,3.5$ obtained from the numerical solutions of the BP equations on single instances (points) and at the ensemble level (lines).}
\label{fig.2}
\end{figure}

\section{Results in the  BP Approximation}
The existence of a topology induced inverse behavior is confirmed by the more 
accurate Bethe-Peierls (BP) approximation \cite{2}. 
On a tree-like structure emerging from a randomly chosen node $i$, the partition function of the
system can be written as
\begin{equation}
Z = \sum_{s_i} e^{-\beta \Delta s_i^2} \prod_{j \in \partial_i} m_{ij}(s_i)
\end{equation}
where $\partial_i$ is the set of neighbors of $i$ and $m_{ij}(s_i)$  
satisfies the recursive equation
\begin{equation}
m_{ij}(s_i) = \sum_{s_j} e^{\beta(s_i s_j -\Delta s_j^2)} \prod_{k \in \partial_j \setminus i} m_{jk} (s_j).
\end{equation}
With the parametrization $m_{ij} (s_i)= A_{ij} e^{\beta( u_{ij} s_i- v_{ij} s_i^2)} $,
 we can easily obtain a set of recursive equations for the {\em fields} $u_{ij}$ and the
{\em anisotropies} $v_{ij}$,
that can be solved numerically and
used to compute the local densities $p_i=\langle s_i^2 \rangle$ and local  magnetization $m_i=\langle s_i \rangle$ for each node $i$:
\begin{eqnarray}
p_i &=& \frac{2\cosh(\beta \sum_j u_{ij})}{e^{\beta(\Delta + \sum_j v_{ij})}
+2 \cosh(\beta \sum_j u_{ij})}, \\ \nonumber
m_i &=& p_i \tanh(\beta \sum_j u_{ij})).
\end{eqnarray}

In Fig.\ref{fig.2} (left) we plot the total magnetization 
as a function of the temperature
$T$ in random graphs with $P (k) \propto k^{-\gamma}$ ($N = 10^4$ ). For all values of $\gamma$,  
the magnetization at $T = 0$ drops to zero for sufficiently large $\Delta$, even if a metastable ferromagnetic
phase survives at $T = 0$. The agreement between BP
results (lines) and Monte Carlo simulations (points) is very good.
In the right panel of Fig.\ref{fig.2}, we report the phase diagram
obtained using the BP approximation for the same values of $\gamma$. The points are obtained averaging the results obtained running BP equations on given
instances, while the full line is the analytic prediction of the continuous transitions and the tricritical points obtained again within the BP approximation, but solving the equations at the level of ensembles
of random graphs. Indeed, for random graphs characterized only by the degree distribution $P(k)$ 
we can divide the nodes in classes according to their degree $k$ and consider the cavity fields $\{u_k,v_k\} $ symmetric within each class. Once introduced the average cavity fields 
$\bar{u}=\sum_k P(k) u_k$ and $\bar{v}=\sum_k P(k) v_k$, one can easily check that they have to satisfy a closed set of self-consistent equations,  $\bar{u} = f(\bar{u},\bar{v})$, $\bar{v} = g(\bar{u},\bar{v})$.
The paramagnetic solution of these equations reads 
\begin{eqnarray}
&&\bar{u}=0 \nonumber\\ 
&&\bar{v}^* =  \frac{1}{\beta}{\sum_k \frac{k P(k)}{\langle k \rangle} \log \frac{e^{\beta (\Delta+(k-1)\bar{v}^*)}+2}{e^{\beta(\Delta+(k-1)\bar{v}^*)}+2\cosh\beta}}. 
\end{eqnarray}
Upon expansion around this solution, the tricritical point is given by the conditions 
$\partial_{\bar{u}}f(0,\bar{v}^*)=1$, $\partial_{\bar{u}}^3f(0,\bar{v}^*)=0$.
The first equation describes the $\lambda$ line of the second order critical points and gives us
the value of $\Delta_c(\beta)$ as a function of the temperature while the second equation fixes the ending point of this line,  
$\beta_c$ and its associated $\Delta_c(\beta_c)$. The previous conditions result into a set of two coupled implicit equations for $\beta_c$ and $\Delta_c$
\begin{eqnarray}
\sum_k  \frac{k(k-1)P(k)}{\langle k \rangle} \frac{2\sinh\beta_c}{e^{\ell_c}+2\cosh\beta_c} &= &1\nonumber\\
\sum_k  \frac{k(k-1)^3 P(k) }{\langle k\rangle}\frac{e^{2\ell_c}-2e^{\ell_c}\cosh\beta_c -8}{\left(e^{\ell_c}+2\cosh\beta_c\right)^3}&=&0
\end{eqnarray}
where $e^{\ell_c}=e^{\beta_c(\Delta_c+(k-1)\bar{v}^*)}$
These equations have been solved numerically for ensembles of power-law random graphs with $\gamma=2.5,3,3.5$ (see right panel in Fig. \ref{fig.2}).

The overall picture obtained from the CW approximation is confirmed by BP results, although the position of the tricritical
point seems to move to the left of the maximum
of the curve in the BP phase diagram, showing that
inverse transitions can be continuous  as well. 

\begin{figure}
\onefigure[width=0.49\textwidth]{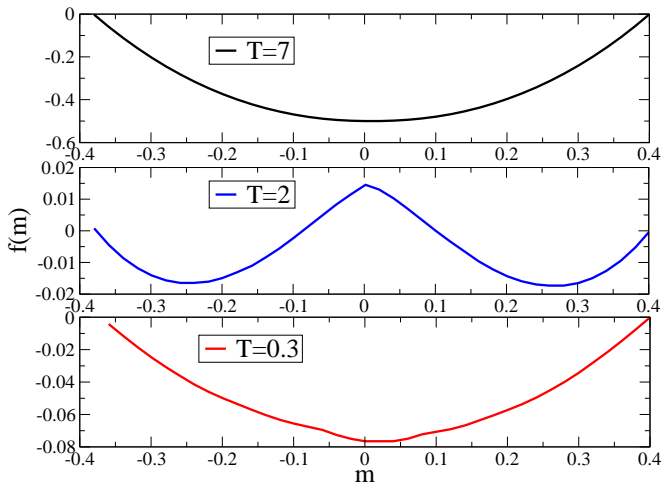}
\caption{The free energy computed on a single graph is shown for $\Delta=4$ in the left panel and for $\Delta=3$ in right panel.  We use the algorithm developed in \cite{21}, by using a Legendre transformation. The scale free graph used has $N=5000$, $k_{min}=4$ and $\gamma=3.5$. In both cases the free energy displays a minimum in zero for $T>T_c$ the critical temperature and, more importantly, for $T<T_m$ (the melting temperature). In between these two temperatures, $T_m<T<T_c$, it shows a ferromagnetic behavior with two minima in $\pm m$. }
\label{fig:freeeenergy}
\end{figure}

We checked that the fixed point solutions of the cavity equations have a true thermodynamic meaning by computing
and plotting the Landau free energy as a function of the magnetization  $f(m)$. 
We use the algorithm developed in \cite{21} to reconstruct
the free energy landscape by means of a Legendre transformation.
The two panels reported in Fig.\ref{fig:freeeenergy} show that the free energy of a scale free graph for $\Delta=4$ (left) and $\Delta=3$ (right). We can see that 
in both cases the free energy signals the presence of a stable paramagnetic solution at high temperature, i.e. it has a  minimum at $m=0$. Then a ferromagnetic solution $m>0$ appears at smaller temperature and finally, 
cooling down the system below the melting temperature $T_m$, it is possible to see the minimum 
of free energy again at the origin $m\approx 0$. 


\begin{figure}
\onefigure[width=0.49\textwidth]{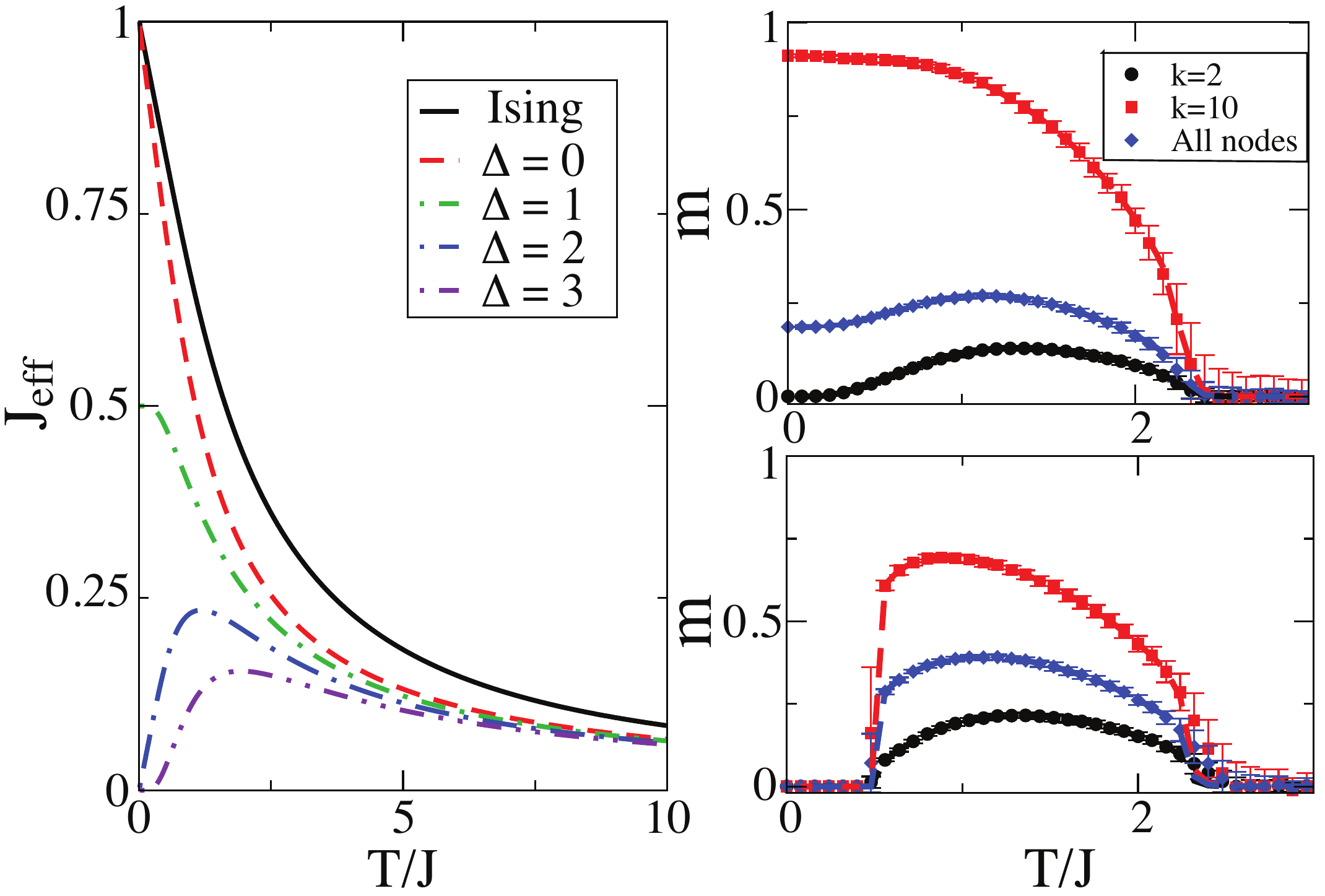}
\caption{Left) 
Effective interaction $J_{eff}(T)$ 
between the neighbors of a node of degree $2$ 
as a function of the temperature $T$
for several $\Delta$.
Right) Magnetization as a function of the temperature for
$\Delta = 3$ in uncorrelated (top) and disassortative (bottom) bi-
modal random graphs with $P(k) = 0.2\delta_{k,2} + 0.8\delta{k,10}$. 
The behavior of nodes belonging to different classes of degrees is
highlighted. The curves display results from both Monte Carlo
simulations (points) and BP calculations (solid lines).}
\label{fig.3}
\end{figure}

\section{A Microscopic Mechanism for Reentrance}
By performing a low temperature analysis, we
found that nodes with different degrees can have
very different values of magnetization.  
In Fig.\ref{fig.3}(top-right) we report the magnetization
curves $m(T)$ at $\Delta = 3$ for a bimodal random graph with
$P(k) = 0.2\delta_{k,10} + 0.8\delta_{k,2}$ obtained by Monte Carlo simulations and BP calculations. 
The curve referring to nodes with lower degree 
is showing a reentrant behavior, whereas the one of high degree nodes is not.
This behavior suggests that the topology-induced
reentrance could be activated at low $T$ by a complex
recursive microscopic mechanism starting from the spins of lowest
degree that freeze to zero for a sufficiently large chemical
potential $\Delta$. For simplicity we consider a node $i$ of degree
$2$ and two neighboring nodes $i_1$, $i_2$ of degree larger than
$2$. At zero-temperature a chemical potential $\Delta > 2$ 
is sufficient to freeze
$s_i = 0$ independently of the neighbors of $i$. Once $s_i$ is
frozen to zero, the two neighbors  $i_1$, $i_2$ are not directly
correlated anymore, the effective interaction is zero and the local fields acting on them do decrease. 
They now have to be compared
with $\Delta$ to establish whether nodes $i_1$, $i_2$
are set to zero or not. Pushing forward this decimation
procedure (which coincides with $\Delta$-core percolation \cite{16})
makes it possible to recursively determine the whole
ground state of the system. We can also check how
the interaction between $s_{i_1}$ and $s_{i_2}$ changes at finite $T$.
If we sum over the values of the internal spin $s_i$, the
partial partition function of the three spins becomes
that of a system of two spins with an effective coupling
$J_{eff} = \frac{1}{2\beta} \log\left(\frac{1+2 e^{-\beta \Delta}\cosh(2 \beta)}{1+2 e^{-\beta \Delta}}\right)$. 
In the left part of Fig. \ref{fig.3}
we show that as soon as $ \Delta> 1$, 
the coupling between $i_1$ and $i_2$ becomes non-monotonic in $T$ and for
$\Delta > 2$ the coupling vanishes at $T = 0$ due to the freezing
of $s_i = 0$. Combining the renormalization argument with
the previous decimation algorithm, it is possible to check
that two far-away spins remain completely uncorrelated
at low $T$ and their effective coupling increases at larger
temperatures. Degree heterogeneity is the first essential
ingredient needed for an inverse transition because when
all nodes have approximately the same degree, the
renormalization of couplings is not effective. Another important ingredient that enhances the reentrant behavior are degree anticorrelations.
In order to verify this idea, we used a Monte Carlo algorithm proposed in\cite{18} in order to generate instances of correlated networks where high degree nodes
are preferentially connected to low degree ones. An uncorrelated
random graph is first generated with the configuration
model. Then Monte Carlo moves are used to exchange
the ends of randomly chosen link-pairs, in order to
introduce positive or negative correlations among the
degrees without altering the degree distribution. In the
bottom-right panel of Fig.\ref{fig.3} we show the curves $m(T)$ for
the BC model in a bimodal random graph with disassortative mixing. In this case the freezing of low degree
nodes at low $T$ is sufficient to disconnect the whole graph
triggering an inverse phase transition. Similarly, a low
level of disassortative mixing is sufficient to enhance the reentrant 
 behavior in scale-free graphs (not shown). Therefore, disassortative mixing favors
inverse phase transitions. Unlike mean-field results,
this argument is not affected by spatial fluctuations,
hence we guess that a topologically induced inverse
transition should be observable in dimensional lattices as
well if there is a sufficient degree of local disorder and
degree anti-correlation. 
The renormalization argument
as well as the mean-field calculation can be extended to
other tricritical spin models, such as the random-field
Ising model (RFIM) with a bimodal distribution of fields.
However, the phase diagram of this model has a
complex structure of singularities even on homogeneous
random graphs \cite{20} and further investigations are needed
to assess the generality of the observed phenomenon.
\section{Conclusions} 
We showed that tricritical spin systems can admit inverse phase transitions on heterogenous graphs. 
Mean-field arguments qualitatively suggest a different scaling of the first and second order critical point with the moments of the degree distribution for random graphs. 
A microscopic mechanism of coupling renormalization shows that in the presence of degree heterogeneity, 
the effective coupling among two nodes becomes temperature dependent. 
In this way, we have shown that the conditions usually described in the Flory-Huggins theory of inverse melting can have also a topological nature. 
The reentrant behavior is strongly enhanced (damped) by the presence of negative (positive) degree correlations.

\acknowledgments
DDM warmly thanks Luca Leuzzi for fruitful discussions and useful suggestions.

\end{document}